\documentclass[%
 amsmath,amssymb,
 aps,
pre, twocolumn
]{revtex4-2}

\usepackage{graphicx}% Include figure files
\usepackage{dcolumn}% Align table columns on decimal point
\usepackage{bm}% bold math
\newcommand{\xv}{{\bf x}}
\newcommand{\pv}{{\bf p}}
\newcommand{\Qt}{{\bf Q}}
\newcommand{\nv}{{\bf n}}
\newcommand{\mv}{{\bf m}}
\newcommand{\lv}{{\bf l}}
\newcommand{\grad}{{\bf \nabla}}

\begin{document}

\title{Tracing chirality from molecular organization to triply-periodic network assemblies: \\ 
Threading biaxial twist through block copolymer gyroids
}% Force line breaks with \\

\author{Ishan Prasad}
\affiliation{Department of Chemical Engineering, University of Massachusetts, Amherst, Massachusetts 01003, USA}%Lines break automatically or can be forced with \\
\author{Abhiram Reddy}%
\affiliation{%
 Department of Polymer Science and Engineering, University of Massachusetts, Amherst, Massachusetts 01003, USA 
}%

\author{Gregory M. Grason}%
\affiliation{%
 Department of Polymer Science and Engineering, University of Massachusetts, Amherst, Massachusetts 01003, USA
}%

 \email{grason@umass.edu}

\date{\today}% It is always \today, today,

\begin{abstract}
Chirality transfer from the level of molecular structure up to mesoscopic lengthscales of supramolecular morphologies is a broad and persistent theme in self-assembled soft materials, from  biological to synthetic matter.  Here, we analyze the mechanism of chirality transfer in a prototypical self-assembly system, block copolymers (BCPs), in particular, its impact on one of the most complex and functionally vital phases: the cubic, triply-periodic, gyroid network.  Motivated by recent experimental studies, we consider a self-consistent field model of ABC* triblock copolymers possessing an end-block of chain chemistry and examine the interplay between chirality at the scale of networks, in alternating double network phases, and the patterns of segmental order within tubular network domains.  We show that while segments in gyroids exhibit twist in both polar and nematic segmental order parameters, the magnitude of net nematic twist is generically much larger than polar twist, and more surprising, {\it reverses handedness} relative to the sense of polar order as well as the sense of dihedral twist of the network. Careful analysis of the intra-domain nematic order reveals that this unique chirality transfer mechanism relies on the strongly biaxial nature of segmental order in BCP networks and relates the {\it biaxial twist} to complex patterns of frame rotation of the principal directors in the intra-domain texture.  Finally, we show that this mechanism of twist reversal leads to chirality selection of alternating gyroid networks in ABC* triblocks, in the limit of very weak chirality . 

\end{abstract}

\maketitle

\section{Introduction}

Chirality transfer, from local interactions between constituents to complex and self-organized patterns of order on larger length scales, pervades material systems~\cite{Harris1999, Efrati2014}, from chiral magnets~\cite{Cheong2022} to biological~\cite{Neville, Selinger2001, Bouligand2008, Sharma2009} and synthetic~\cite{Goodby1991} self-assemblies. Perhaps the simplest example is the cholesteric phase~\cite{Kamien1997, Pieraccini2011}, where non-centrosymmetric interactions of anisotropic subunits, such as rod-like mesogens or biopolymers, stabilize states of mean alignment in which a molecular director rotates helically along an axis perpendicular to itself.  Remarkably, the templating to orientation order (handedness and pitch of this rotation) is manifested at length scales several orders of magnitude larger than those molecular constituents.  In self-organized states with both positional and orientational order, like twist-grain boundary (SmC*)~\cite{Lubensky1990, Goodby2002}, helical nanofilament~\cite{Hough2009, Matsumoto2009} and blue phases~\cite{Wright1989}, the mechanisms and outcomes of chirality transfer are often far more complex owing to additional levels of hierarchy introduced in multiple, coupled order parameters. 

In this article we describe the mechanisms of chirality transfer to triply periodic network phases of soft, self-assembling molecules, focusing on a particular example of block copolymer (BCP) melts~\cite{Bates1990}.  On one hand, BCPs provide a prototypical system for understanding self-assembly in a much broader class of supramolecular systems~\cite{Matsen2002, Reddy2021}.  Beyond this, the chemical versatility of BCPs~\cite{Bates2012, Polymeropoulos2017} enables numerous routes to engineering nanostructured materials with controlled functions that rely on their complex self-assembled structures.

Even in the absence of intrinsic chirality, triply-periodic networks~\cite{Thomas1988, HYDE1997} -- like the double-gyroid (DG), double-diamond (DD), or double-primitive (DP) phases -- are among the most structurally intricate and functionally vital states that form in BCPs and soft supramolecular systems more broadly.  Grossly speaking, these phases are composed of two networks of tubular domains that are divided by a slab-like matrix domain~\cite{Hajduk1994, Schroder-Turk2006}, the undulating shape of which closely approximates a finite-thickness variant of the associated triply periodic minimal surface~\cite{Schoen2012} (e.g. G or D minimal surfaces, respectively).  The tubular domains -- of distinct chemical composition from the matrix region -- connect in multi-valent junctions (e.g. 3- and 4-valent connections), with each single constituent network interlinking with its partner to form an intercatened double network~\cite{Reddy2021, Olmsted1998}.  The unique combination of nanoscopic dimensions, polycontinuous topology, and triply periodic symmetries of self-assembled double networks imbue them with remarkable functionality\cite{Lee2014}, from photonic architectures appearing in birds, beetles, and butterflies~\cite{Saranathan2010, Saranathan2021}, to hybrid plasmonic or photonic metamaterials~\cite{Ullal2004, Hur2011}.  Notably the most exotic of these targeted properties, such as topologically-protected wave propagation and negative refraction, require control over the centrosymmetry of the networks, in other words, they require chirality at the network scale~\cite{Oh2013, Lee2022, Fruchart2018}.

\begin{figure}
% \centering\includegraphics[width=1.0\linewidth]{fig_monomer_design.eps} 
\centering\includegraphics[width=1.0\linewidth]{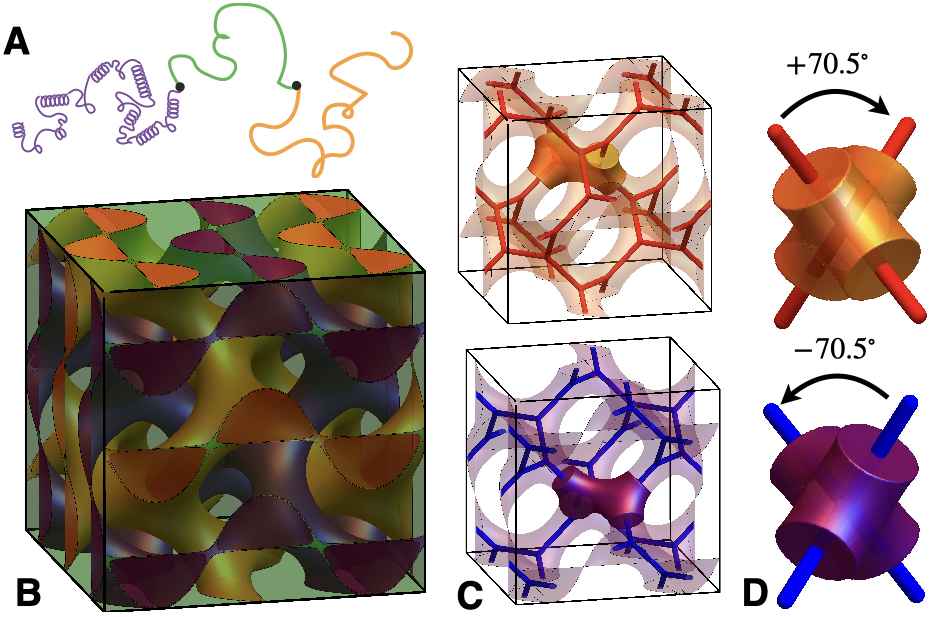} 
\caption{ {\bf Chirality selection in alternating double-gyroids of ABC* triblock copolymers.} (A) A schematic illustration of ABC* copolymer possessing a chiral A (purple) block and two achiral B (green) and C (gold) blocks.  (B) shows alternating double gyroid (aDG) with domains colored according to end blocks composing the chiral single gyroid networks and mid-block composting the slab-like matrix.  (C-D) illustrate the mesoscopic chirality of the single-gyroids, with skeletal graphs the networks highlighted as red and blue, repsective, from right- and left-handed chiral.  The network chirality can be associated with handedness of the dihedral twist between two adjoining 3-valent nodes of the network, as shown in (D).}
\label{fig:aDG}
\end{figure}

Among the cubic double-network phases, gyroid networks possess a unique capacity for chirality, in that each constituent single-gyroid network is chiral~\cite{Foerster1994,Schick1998}.  However, in the simplest case, i.e. linear AB diblock where the composition of each tubular network is the same (say, A-type), the stable DG phase is achiral with two enantiomeric single-gyroid domains arranged in an inversion-symmetric fashion according to the $Ia\bar{3}d$ symmetry~\cite{Hajduk1994}.  Introducing more components, such as linear ABC triblocks, opens up the possibility of breaking this ``exchange symmetry'' between network domains, leading to {\it alternating} double networks where one tubular network domain is A-type, its complement is C-type, both of which are divided by a B-type matrix~\cite{Epps2004}. Indeed, experimental~\cite{Epps2004B} and theoretical~\cite{Tyler2007, Qin2010} studies of ABC melts show windows of alternating DG (aDG), the morphology shown in Fig.~\ref{fig:aDG}, the chirality of this structure can be identified as right- or left-handed, according to the sense of dihedral rotation between the 3-fold axes of adjacent nodes in a single network domain~\cite{Prasad2018, Feng2019}.  Although each tubular domain of the aDG is chiral at the mesoscopic network scale(with space group $I4_1 3 2$) nearly all triblock systems are {\it achiral} at the molecular level. Hence, for achiral BCP the chirality of the phase that forms is determined {\it randomly} via spontaneous symmetry breaking~\cite{Hur2011, Vignolini2012}, limiting possibilities to exploit the functional utility of the chiral variant of this complex nanostructure.  

Recent experiments by Wang et al.~\cite{Wang2020} give evidence that the left- vs. right- symmetry is broken in triblocks possessing a chiral A- block (which we refer to as ABC* triblocks), in their case stereopure polylactide (PLLA). Depending on whether the polylactides are D- or L-type, the P(D/L)LA blocks have a preference to compose the single-gyroid domain of a particular handedness, in effect selecting the net chirality of the formed aDG from the chirality of the molecular constituents.  This observation raises a basic question: How does the left- vs. right-gyroid network symmetry couple to the organization of chiral chain segments within those domains?  

Prior studies of chiral diblocks have shown the ability to transfer the chirality of a chain backbone to columnar phases, known as helical cylinder H* phase~\cite{Ho2009,Wang2019}.  These observations have been rationalized by the orientational self-consistent field (oSCF) theory of chiral diblocks, that couples the statistical confirmations of random walking BCPs with gradient free energy of chiral segments within nanophase-separated domains~\cite{Zhao2012, Zhao2013}.  This oSCF theory links the mesoscopic helicity of H* domains with the threading of handed cholesteric segment twist within the tubular domains of the chiral block, provided that the preference for chiral twist of those segments (i.e. the preferred inverse cholesteric pitch) is sufficiently strong~\cite{Grason2015}.  

In this article, we describe the mechanism of chirality transfer from the segment-scale to the mesoscale network chirality in this class of chiral ABC BCP based on the orientational self-consistent field  (oSCF) theory.  We show that segmental twist is most pronounced in the (tensorial) nematic order, and surprisingly, reverses the handedness of not only the polar segmental order but also the mesoscopic twist of the single-gyroid domain itself.  We show that this chirality reversal derives from unique features of twist in a {\it biaxial} order parameter field.  Finally, we show that the thermodynamics of the subdomain pattern of twist has the ability to select the equilibrium chirality of aDG phases of ABC* triblocks, even in the limit of weakly-anisotropic segmental interactions.

\section{Self-consistent field theory of intra-domain segment texture}

We begin by analyzing the intrinsic patterns of segmental order in alternating networks of ABC triblocks with isotropic interactions (see Appendix~\ref{sec: oSCF} for details of oSCF theory). In this limit, the enthalpy of segment interactions is described by local coupling of purely {\it scalar} composition fields, $\phi_\alpha( \xv)$, the volume fraction of type-$\alpha$(=A,B,C) segments at $\xv$, with a mean-field free energy of the standard Gaussian chain BCP form~\cite{Tyler2007},
\begin{multline}
\label{eq: Fiso}
    \frac{F_{\rm iso}}{k_B T} = \rho_0 \int d^3 \xv\bigg\{\chi_{AB} \phi_A\phi_B  + \chi_{BC} \phi_B  \phi_C  + \chi_{AC} \phi_A  \phi_C  \bigg\} \\  - \frac{\rho_0 V}{N} ~S \big[\phi_A , \phi_B , \phi_C \big] ,
\end{multline}
where $\rho_0$ is the segment density, $N$ is the total segment number in chain, $V$ is the total volume and $\chi_{\alpha \beta}$ is the Flory-Huggins parameter describing (scalar) interactions between $\alpha$- and $\beta$-type segments. The inhomogeneous composition profiles in combination with the incompressibility constraint, $\sum_\alpha \phi_\alpha (\xv) =1$, lead self-consistently to spatially varying chemical potentials for segments that reduce the random-walk entropy of the total $\rho_0 V/N$ chains in the system, encoded in the mean-field entropy functional $S\big[\phi_A , \phi_B , \phi_C \big] $.  We consider a melt of ABC triblocks composed of $N_\alpha = f_\alpha N $ $\alpha$-type segments, where $N$ is the total chain length.  We focus mainly on the case of symmetric end blocks $f_A=f_C=(1-f_B)/2$, equal segment lengths $a$ in each block, and the possibility of segmental chirality only in the A block.   For given (scalar) interaction parameters and block lengths, self-consistent solutions of this isotropic model, eq. (\ref{eq: Fiso}), solve the equilibrium composition profiles as well as the end-distribution functions $q^{\pm}(\xv,n)$, which gives the partial statistical weight of random-walk chain segments diffusing from its respective A or C free ends to the $n$ segment at position $\xv$, from which the mean-field partition function of the entire chain is ${\cal Z} = \int d^3\xv ~q^{+}(\xv,n)q^{-}(\xv,n)$~\cite{Matsen2002}.  End-distributions determine not only the scalar composition profiles, $\phi_\alpha(\xv) = \frac{V}{N {\cal Z}} \int_{n\in \alpha} dn ~ q^{+}(\xv,n)q^{-}(\xv,n)$, but also mean-field orientational order parameters~\cite{Prasad2017,Burke2018}:  the (vectorial) polar order, 
\begin{equation}
    {\bf p}_\alpha (\xv) = \frac{V a}{N {\cal Z}} \int_{n\in \alpha} dn ~{\bf J} (n,\xv)
\end{equation}
where ${\bf J} (n,\xv) =  (q^+ \grad q^- - q^- \grad q^+ )/6$; and the (tensorial) nematic order,
\begin{equation}
    {\bf Q}_\alpha (\xv) = \frac{Va^2}{N {\cal Z}} \int_{n\in \alpha} dn \bigg({\bf K} (n,\xv) - \frac{{\bf I}}{3} {\rm Tr}\big[{\bf K}(n,\xv)\big] \bigg),
\end{equation}
with the tensor elements $K_{ij}=\big(q^+ \partial_i \partial_j q^-+q^- \partial_i \partial_j q^+ - \partial_i q^+ \partial_j q^- - \partial_i q^- \partial_j q^+\big)/60$, where ${\bf I}$ is the identity and $i,j,k$ are spatial indices~\footnote{Here, we use the result for the large-$N$ (Gaussian) limit of the freely-jointed chain model, which gives identical results for $\phi$ and ${\bf p}$ as the Edwards chain (a.k.a. Gaussian thread) model and only a different numerical prefactor for ${\bf Q}$~\cite{Prasad2017}}.  Notably, spatially modulated phases imply anisotropy in the chain configurations, hence all ordered phases imply non-vanishing patterns of both polar {\it and} nematic order in the underlying segments, even in the absence of explicit segmental anisotropy.  While $ {\bf p}_\alpha (\xv) $ tracks the orientation of segments by distinguishing between the ``head to tail'' orientation of segments along the chain, the tensorial order parameter $ {\bf Q}_\alpha (\xv) $ tracks only the axis of segment alignment, as is familiar to molecular theories of liquid crystallinity~\cite{deGennesProst}.

However, because these order parameters encode distinct orientational symmetries (i.e. head-to-tail vs. axial orientation) they exhibit quite distinct spatial patterns and magnitudes, even in the same ordered morphology~\cite{Prasad2017}.  Crudely speaking, the polar order tends to splay from the center of domains, roughly normally to the inter-material dividing surface (IMDS) between them.  Nematic order is far more complex, showing a narrow zone of tangential alignment at the IMDS as well as a director that shows predominantly normal alignment deep in the brush zone.  Even more significant is that intrinsic nematic order becomes {\it biaxial} in curved morphologies, exhibiting a secondary (minor) director field along the axis of convex curvature, in addition to principal normal alignment of the brush-like domains.  Given the coexistence of both types of order and the distinctions between their patterns, it is not {\it a priori} clear whether the thermodynamically relevant measure of segment chirality derives from polar or nematic order.  

\begin{figure*}
% \centering\includegraphics[width=1.0\linewidth]{fig_monomer_design.eps} 
\centering\includegraphics[width=0.8\linewidth]{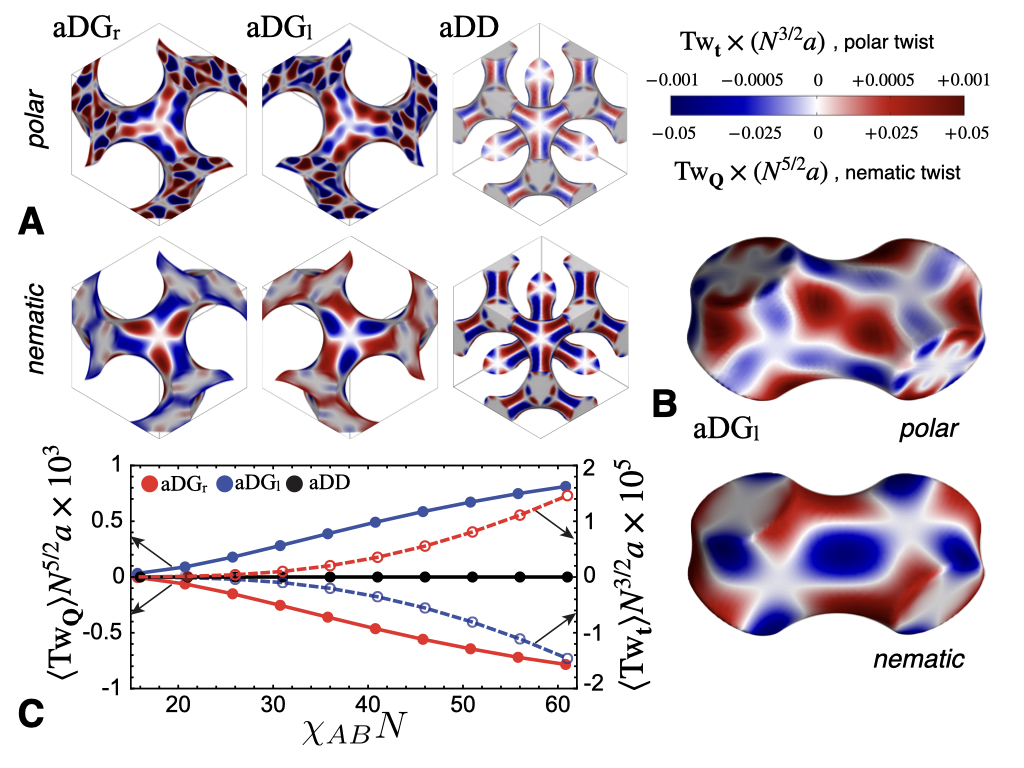} 
\caption{{\bf Intradomain twist in alternating double networks.} (A) shows maps of the local (polar and nematic) segment twist density in tubular network domains for $\chi_{AB}N =\chi_{BC}N =0.37\chi_{AC}N=41 $ and $f_A =f_C= 0.24$:  A-component forming a right-handed network of alternating gyroid (aDG$_{\rm r}$); A-component forming a left-handed network of alternating gyroid (aDG$_{\rm l}$); and A-component forming a 4-valent network of alternating double diamond (aDD).  (B) highlights the difference in spatial patterns of polar vs. nematic twist on a 2-node region of the aDG$_{\rm l}$ structure.  Notably, the contours of locally right-/left-handed segment {\it polar} twist follow the dihedral rotation between nodes, while the corresponding patterns of {\it nematic} twist wind counter to the node twist. (C) plots the mean segment twist of A-domains in aDG$_{\rm r}$, aDG$_{\rm l}$ and aDD versus segregation strength for $\chi_{AB}N =\chi_{BC}N =0.37\chi_{AC}N$ and $f_A =f_C= 0.24$.  While aDD has regions of local segment twist, as an achiral morphology, the mean twist is zero.  In contrast, mean polar and nematic twist are non-zero for gyroidal domains, but of distinct magnitudes and signs.  }
\label{fig:intradomain}
\end{figure*}

\section{Intrinsic twist of alternating network domains: polar vs. nematic}

We compare the twist of both polar and nematic order parameters, of the A end block, defined in the case of liquid crystals as the lowest-order pseudo-scalar gradients of $\pv_A$ and $\Qt_A$~\cite{Wright1989},
\begin{equation}
    {\rm Tw}_\pv \equiv \pv_A \cdot (\grad \times \pv_A) ; \ \ {\rm Tw}_\Qt \equiv \Qt_A \cdot (\grad \times \Qt_A),
\end{equation}
For the most commonly studied chiral texture, the uniaxial cholesteric, both ${\rm Tw}_\pv$ and ${\rm Tw}_\Qt$ have a simple interpretation -- the (+/-) sign indicates (right/left) handedness of the director winding and the magnitude is inversely proportional to the pitch~\cite{Harris1999}.  Here, we focus on the twist of end-block, A, segments composing a tubular network domain, but in Appendix~\ref{sec: intrinsic} we analyze an example of twist in matrix domains, as well as computational details.

In Fig.~\ref{fig:intradomain}A-B, we compare the distributions of polar and nematic twist of the A-domain in three SCF-predicted alternating network phases of ABC triblocks, right- and left-handed aDG (denoted as aDG$_{\rm r}$ and aDG$_{\rm l}$) and the alternating double-diamond (aDD), for the common parameters $\chi_{AB}N =\chi_{BC}N =0.37\chi_{AC}N=41 $ and $f_A =f_C= 0.24$.  First, we note from the spatial maps (projected on the AB IMDS) that there is a coexistence of regions of positive and negative twist (both polar and nematic) for all networks.  As the aDG$_{\rm r}$ and aDG$_{\rm l}$ are related by an inversion symmetry, local regions of positive twist on one structure map onto regions of negative twist on the corresponding (inverted) enantiomeric network domain.  

Fig.~\ref{fig:intradomain}C plots the spatial average of the A segments of both types of twist as a function of segregation strength for all three networks.  First, we note that there is zero net twist in aDD, which derives from the fact that as a mirror symmetric structure, $Fd\bar{3}m$, any local ``patch'' of positive segment twist maps onto an equivalent patch of negative segment twist.  For the chiral aDG domains, on the other hand, positive and negative twist do not balance, and each domain includes an enantiomeric excess of one sign of twist.  The net {\it polar} twist, which increases with inter-block segregation strength, is equal and opposite on the two aDG domains following the same sign of twist as the dihedral twist itself (i.e. $\langle {\rm Tw}_\pv \rangle <0$ in the left-handed network domain and vice versa).

The nematic twist is similarly non-zero and increasing with segregation strength for the chiral aDG domains. However, it exhibits two key differences from the polar twist.  First, the magnitude of the mean twist, when rescaling for the intrinsic dependencies on the characteristic chain dimension $a N^{1/2}$, is significantly greater for nematic order than for polar order.  Second, the net {\it sign} of nematic twist is reversed relative to the polar twist as well as relative to the sense of dihedral rotation of the gyroid network itself (i.e. $\langle {\rm Tw}_\Qt \rangle >0$ in the left-handed network domain and vice versa).

\begin{figure*}
% \centering\includegraphics[width=1.0\linewidth]{fig_monomer_design.eps} 
\centering\includegraphics[width=0.8\linewidth]{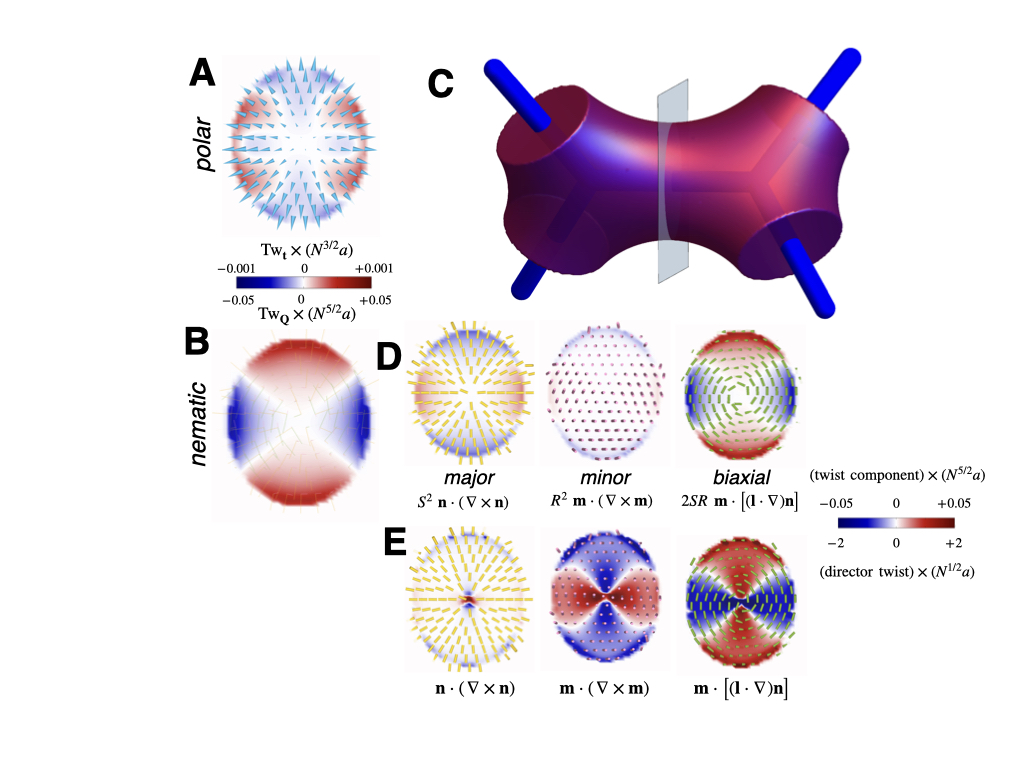} 
\caption{ {\bf Subdomain patterns of network twist}.  Heat maps of local segment twist density are shown for polar twist (A) and nematic twist (B) in the cross-section of the A-segment gyroid network of the aDG$_{\rm l}$ phase, for a the planar cut normal to the strut, as illustrated in (C).  Arrows indicate the local polar orientation in (A), while principal directions of nematic tensor are highlighted by yellow, green and blue lines in (B).  In (D), the components of nematic twist, defined in eq. (\ref{eq: twistdecomp}), are mapped to the strut cross-section.  In (E), the corresponding maps of major director twist, minor director twist and biaxial frame twist, are shown for the strut section.}
\label{fig:decomposition}
\end{figure*}

To understand the stronger magnitude and reversed sense of nematic twist relative to polar, it is illustrative to consider the twist textures in cross-sections of the A-block domain strut of aDG$_{\rm l}$ (the planar cut shown in Fig.~\ref{fig:decomposition}C).  First, we note that polar order (Fig.~\ref{fig:decomposition}A) tends to orient radially away from the strut axis in Appendix~\ref{sec: aligment}. To be more precise, as described previously~\cite{Prasad2017} and analyzed in Appendix~\ref{sec: aligment}, polar order predominantly follows the gradients of segment composition, so that to a first approximate $\pv_A \propto  - \grad \phi_A$, which can be expected purely on symmetry grounds for sufficiently weak segregation, and broadly due to the extension of chain trajectories normal to the IMDS.  Because $\grad \times (\grad \phi_A) = 0$, the polar twist is only non-zero if the polar director deviates from $-\grad \phi_A$.  As these deviations are evidently quite small (see Fig.~\ref{fig:alignment}), it follows that the degree of polar twist is vanishingly small in aDG domains, as well as for any other ordered BCP domain structure, in the absence of orientational segment interactions.

Fig.~\ref{fig:decomposition}B shows the the spatial pattern of nematic twist in aDG$_{\rm l}$ domain, for comparison. We note again the larger magnitude of scaled nematic twist, but also, and more striking, the spatial regions of local left- vs. right-handed twist in the strut cross-section have largely swapped places relative to polar twist.  For example, along the semi-major directions of the elliptical domain sections, the polar twist is negative (i.e. left-handed) following the dihedral twist of the domain.  In contrast, the semi-minor directions of the cross-section exhibit positive (i.e. right-handed) nematic twist.  We next consider the origins of this reversal of segmental twist.

\section{Chirality reversal via biaxial twist}

The nature of nematic twist in aDG domains is far more complex because segmental order is {\it biaxial}. Unlike biaxial nematic phases of mesogens~\cite{Straley1974, Madsen2004}, here biaxiality derives from the anisotropically curved domain shapes realized by BCP packing~\cite{Prasad2017}.  A biaxial order parameter~\cite{Alben1973} can be decomposed into the degree of alignment $S$ along a {\it major axis} $\nv$ and a weaker degree of alignment $0<R<S$ along a {\it minor axis} $\mv \perp \nv$, where these directions correspond to the principal directions of $\Qt_A$ with the largest two eigenvalues,
\begin{equation}
(\Qt_A)_{ij} = S \Big(n_i n_j - \frac{1}{3} \delta_{ij} \Big) + R \Big(m_i m_j - \frac{1}{3} \delta_{ij} \Big).
\end{equation}
Note that both the directors ($\nv$ and $\mv$) and the order parameter magnitudes ($S$ and $R$) can vary spatially according to the directions and degree of segmental alignment. Note also that the first two directors define a third axis $\lv \equiv \nv \times \mv$~\footnote{It is straightforward to show that the eigenvalues of $\Qt_A$ are given by the projections into these principal directions:  $(\Qt_A)_{n n} = (2S - R)/3$; $(\Qt_A)_{m m} = (2R - S)/3$; and $(\Qt_A)_{l l} = -(S+R)/3$.}.  Given this parameterization, the nematic twist can be split into three contributions (see Appendix~\ref{sec: biaxial} for details) ,
\begin{equation}
\label{eq: twistdecomp}
{\rm Tw}_\Qt = S^2 \nv \cdot ( \grad \times \nv) + R^2 \mv \cdot (\grad \times \mv) + 2S R~ \mv \big[ (\lv \cdot \nabla) \nv\big] .
\end{equation}
The first two terms represent the twist of the major and minor director, respectively, while the third term is unique to the biaxial textures (i.e. $S$ and $R$ are non-zero).  We refer to this as {\it biaxial twist}, and it measures the rate of rotation of major and minor directors along the mutually perpendicular co-director $\lv$, as well as the degree of biaxial order.  

\begin{figure*}
% \centering\includegraphics[width=1.0\linewidth]{fig_monomer_design.eps} 
\centering\includegraphics[width=0.8\linewidth]{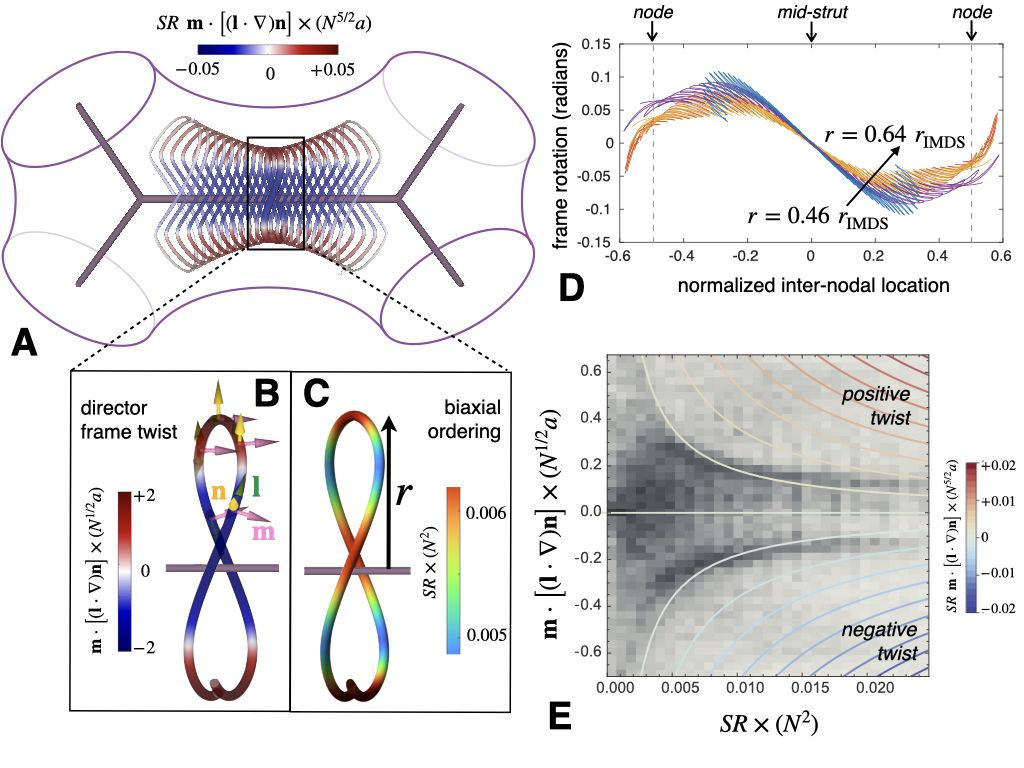} 
\caption{ {\bf Winding of biaxial twist along inter-node gyroid struts}.  (A) shows an integral curve of the co-director $\lv = \nv \times \mv$ computed from the biaxial nematic texture of (left-handed) A domain of aDG$_{\rm l}$, with regions colored red/blue corresponding to local right-/left-handed frame twist.  The geometry of frame-twist is illustrated for a highlighted loop in (B), the major ($\nv$) and minor ($\mv$) directors winding around the $\lv$ along the looping trajectory.  Along this trajectory the degree of biaxiality, as measured by $S \times R$, varies as shown in (C).  (D) plots net rotation of the director frame along the ``orbiting'' integral curves of $\lv$ versus the projected location along the strut (normalized by the node-node separation).  This plot shows multiple curves corresponding to the increasing mid-strut radial distance of the curve from strut $r$, normalized by the distance to the IMDS, $r_{\rm IMDS}$.  (E) gives the 2-point density histogram of values of director frame-twist ($\mv \big[ (\lv \cdot \nabla) \nv\big]$) and biaxiality ($SR$), with colored contours constant biaxial twist ($2 SR \mv \big[ (\lv \cdot \nabla) \nv\big]$) overlaid. }
\label{fig:solenoid}
\end{figure*}

Figure~\ref{fig:decomposition}D-E plots the nematic twist and these three constituent components in the cross-section of the aDG$_{\rm l}$ strut in the core of A domain (i.e. in the region $\phi_A \geq 0.9$).  Notably, the spatial pattern and sign of nematic twist follows closely to the polar twist, consistent with the fact that in the core of the domain, the principal nematic director aligns with the polar order parameter (as shown in Fig.~\ref{fig:alignment}B). Similar tendency of $\nv$ to lock in orientation along composition gradients also implies a relatively weak measure of major twist, consistent with the vanishingly small magnitude of $\nv \cdot (\grad \times \nv)$ in Fig.~\ref{fig:decomposition}E. On the other hand, While there is a reasonably strong degree of minor director twist, $\mv \cdot (\grad \times \mv)$, as shown in Fig.~\ref{fig:decomposition}E, the magnitude of alignment in the minor direction is relatively weak (i.e. $R^2\ll S^2$) and the net contribution from minor twist is negligible.  Finally, the distribution of biaxial twist, in Fig.~\ref{fig:decomposition}D, shows that it is of a substantially larger magnitude and reversed in sense of chirality from major and minor twist.  From Fig.~\ref{fig:decomposition}E, we observe that the {\it frame twist}, $\mv \big[ (\lv \cdot \nabla) \nv\big]$, is comparable in magnitude to the twist of the minor director (with a reversed sense), but as this term is weighted by the product of the magnitude of major and minor axis alignment (i.e. $S \times R$) it dominates the over minor twist.  Because the contribution from biaxial twist dominates over the negligible contributions from the major and minor twist, it sets the net nematic twist and accounts for the reversal of chirality in the nematic order of segments relative to the mesoscopic twist of the aDG networks themselves.

The geometry of biaxial twist can be illustrated in terms of the integral curves of the co-director $\lv$, with the local twist corresponding to the frame rotation of $\nv$ and $\mv$ along these curves.  Fig.~\ref{fig:solenoid}A shows one such curve within an A domain of aDG$_l$, which spans between dihedral nodes and winds, like a solenoid, around the strut.  Along these solenoidal trajectories, the rotation of the $\nv$ and $\mv$ frame experiences both left- and right-handed regions of twist, as highlighted in Fig. \ref{fig:solenoid}B, but notably, the degree of biaxiality, as measured by the product $S \times R$ also varies around this looping ``orbit'' as shown \ref{fig:solenoid}C.  The accumulated rotation of the frame (i.e. integrating the director frame twist along the curves) is plotted in Fig. \ref{fig:solenoid}D, for a series of varying radial distances $r$ (mid-strut) of orbits relative to the strut. Notably, the frame rotation is net negative, consistent with the overall {\it left-handed} writhing of grossly helically winding around the left-handed network strut.  However, the biaxial order in these curves is relatively larger in regions of counter-rotating, right-handed, frame twist, which correspond to the red ``top/bottom" portions of the looping paths in Fig. \ref{fig:solenoid}A-B.  

We note that these counter-twisting regions correspond to the most eccentric regions of the elliptical cross-sections in Fig.~\ref{fig:decomposition}, in other words, regions of highly anisotropic domain curvature.  As described previously~\cite{Prasad2017}, anisotropic domain curvature is directly correlated with degree of biaxial segment order in BCP domains.  Hence, the co-location of counter-twist and enhanced biaxiality leads these ``counter-revolutionary" pockets of segmental order to dominate the net nematic twist in gyroidal networks, that is, they make left-handed network domains overall more favorable to right-handed twist. This bias of strong biaxiality in counter-twisting regions is evident in the 2-point histograms within the A domain of aDG$_l$ showing the frequency of points as a function of both their frame twist, $\mv \big[ (\lv \cdot \nabla) \nv\big]$, and $S \times R$. While points are apparently nearly evenly distributed between right- and left-handed frame twist, right-handed twist regions tend to skew towards larger biaxial order, biasing the product of these quantities (the biaxial twist) to positive (i.e. right-handed) values for a left-handed gyroid network domain. Overall, this analysis reveals how biaxiality, qualitatively and quantitatively, alters the nature of chiral organization, particularly in cases where the principal director is predominantly locked to the normals of iso-density in a SmA-like fashion.

\section{Thermodynamics of alternating networks in ABC* triblocks: weak chirality}

With the analysis for the {\it intrinsic} twist in ABC networks (i.e. with isotropic segment interactions) in mind, we consider the thermodynamic consequences for the coupling of self-assembly to segmental chirality in ABC* triblocks.  The dominance of nematic twist over the polar twist in aDG domains argues that the most significant anisotropic chiral interactions have nematic symmetry, and to a good approximation the effects of potentially polar interaction terms can be neglected. As such, the minimal free energy model includes the lowest gradients in $\Qt_A$ familiar from the $\Qt$-tensor theories of chiral mesogens (e.g. blue phases)~\cite{Wright1989},
\begin{multline}
\label{eq:FQ}
    F^*_{\rm nem}[ \Qt_A] = \frac{1}{4} \int d^3 {\bf x} \bigg\{ K_0 \big[ (\nabla \cdot \Qt_A)_i \big]^2  \\  + K_1  \big[(\nabla \times \Qt_A)_{ij}+2 q_0 (\Qt_A)_{ij} \big]^2\bigg\},
\end{multline}
where $(\nabla \cdot \Qt)_i=\grad_j Q_{ij}$, $(\nabla \times \Qt)_{ij}= \epsilon_{i l k} \grad_l Q_{kj}$ and summation over repeated indices is implied. Here $K_0$ and $K_1$ are elastic constants for divergence and curl of $\Qt_A$, which we set onto an equal value $K = K_0 = K_1$ for this study~\footnote{Elastic constants for the tensor order parameter, $K_0$ and $K_1$, can be related to the standard Frank constants for the nematic director for splay, twist, and bend as well as gradients of $S$, for uniaxial textures.~\cite{Wright1989}}. Segmental chirality, $q_0 \neq 0$, parameterizes the free energy preference for non-zero segment twist:  $q_0 >0$ corresponds to a preference for $\langle Tw_\Qt \rangle <0$, which is a  thermodynamic bias for {\it left-handed} cholesteric twist. Incorporating chiral (nematic) segment interactions leads to a total free energy that combines the (mean-field) enthalpy and entropy of the scalar BCP melt model, eq. (\ref{eq: Fiso}), with the chiral nematic gradient free energy of eq. (\ref{eq:FQ}),
\begin{equation}
F_{\rm tot} = F_{\rm iso} +F^*_{\rm nem} .
\end{equation}
In general, anisotropic segment interactions lead to a self-consistent coupling of random-walk chain configurations and the orientational order parameters, which take the form of self-consistent locally-aligning fields~\cite{Zhao2012, Burke2018}. However, for the purposes of the present study, we restrict our attention to the limit of {\it weakly-anisotropic} segmental interactions. As described in Appendix~\ref{sec:weak}, formally this limit corresponds to a thermodynamic expansion in the elastic constant $K$, where the lowest order term derives simply from computing chiral nematic gradient contribution in eq. (\ref{eq:FQ}) above, from the intrinsic nematic texture, i.e. the biaxial state $\Qt_A (K \to 0)$, analyzed above in Figs.~\ref{fig:intradomain}-~\ref{fig:solenoid}.  

\begin{figure}
% \centering\includegraphics[width=1.0\linewidth]{fig_monomer_design.eps} 
\centering\includegraphics[width=1\linewidth]{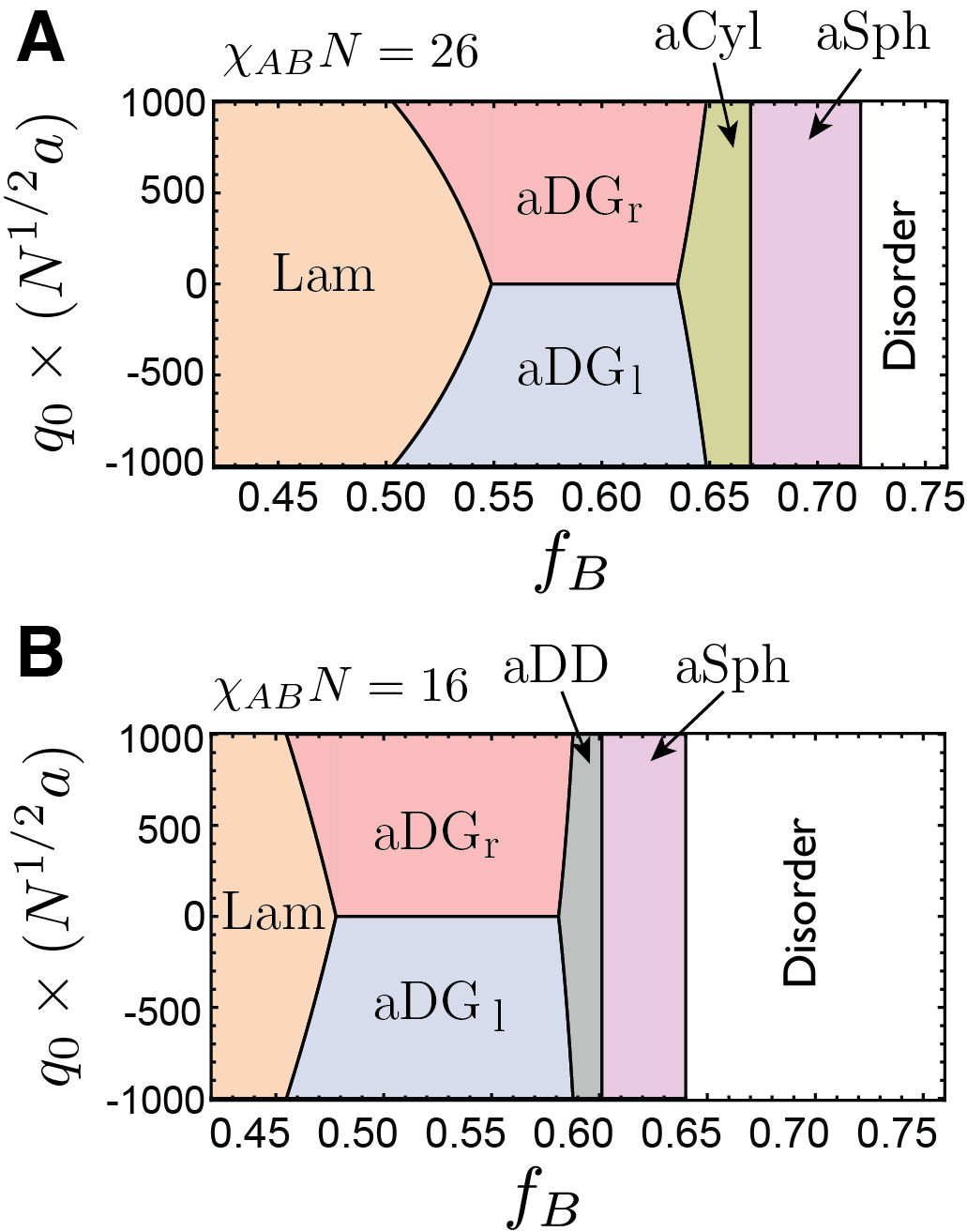} 
\caption{ {\bf Phase behavior of ABC* triblock melts in the limit of weakly anisotropic interactions.} oSCF phase diagrams for symmetric end-block composition ($f_A=f_C$) unfrustrated interactions ($\chi_{AB} N = \chi_{BC} N=0.37\chi_{AC} N$), showing regions of stable morphologies:  lamellar (Lam), alternating (square) cylinder (aCyl), alternating (BCC) spheres (aSph), alternating double-diamond (aDD) and right and left handed alternating double-gyroid (aDG). (A) and (B) correspond to intermediate-segregation ($\chi_{AB} N = 26$) and weak-segregation ($\chi_{AB} N = 16$), respectively.  The strength and handedness of chiral interactions is modulated by $q_0$ the inverse pitch of preferred cholesteric order.  Here, the Frank constants are set to an equal value $K/(\rho_0 a^2 N^2) = k_B T/2$.  }
\label{fig:phase}
\end{figure}

To assess the thermodynamic consequence of intrinsic twist on the assembly of alternating networks of ABC* triblocks, we consider two regions of the large parameter space of ABC triblocks~\cite{Tyler2007,Qin2010}, in particular symmetric end-block composition ($f_A=f_C$), with unfrustrated interactions ($\chi_{AB} N = \chi_{BC} N=0.37\chi_{AC} N$), for relatively weak ($\chi_{AB} N = 16$) and intermediate ($\chi_{AB} N = 26$) segregation strengths, and for a range of effective chiral coupling strengths $q_0$ (see Appendix~\ref{sec:weak}). We first discuss the intermediate segregation, shown in Fig.~\ref{fig:phase}A.  As in the case of purely scalar interactions, studied previously in ref. ~\cite{Qin2010}, with increasing mid-block fraction, we observe a sequence of alternating lamellar (Lam) to aDG - alternating (square) cylinder (aCyl) to alternating (BCC) spheres (aSph) phases.  However, it is important to note that for any $q0 \neq 0$ the left vs. right aDG degeneracy is broken, and again, due to the predominance of biaxial twist in aDG domains, the thermodynamically selected network chirality is of the {\it reversed} handedness with respect to the sense of cholesteric segment twist preferred by same sign of $q_0$.  The expansion of the windows of stable chiral aDG phases relative to achiral competitors with increasing strength of segmental chirality $q_0$, is another notable effect which derives from the unique natural twist of segment packing in gyroidal domains, even in the absence of chirality in constituent segments themselves.  

Fig.~\ref{fig:phase}B shows a similar composition-chirality plane for weaker segregation, as previously reported~\cite{Qin2010}, this regime includes a stability window of aDD intermediate to aDG and aSph.  Again, we note that outward tilt of aDG/aDD phase boundaries with increasing $q_0$. Given the appearance of a triple point between aDG$_{\rm r}$, aDG$_{\rm l}$ and aDD in this phase diagram, it is interesting to note that Wang {\it et al.} observe that PI-PS-PLA with racemic lactic acid blocks form at aDD phase, while the stereopure (L or D type) PLA blocks lead to chiral aDG phases~\cite{Wang2020}.  This suggests that the experimental system may be similarly placed close to a triple point between these three phases, though it remains to be understood how to precisely map the magnitudes of chiral nematic free energy parameters onto this experimental system, and more generally, how the broader structure and stability of alternating diblock phases will vary as segmental interactions become larger. For sufficiently strong anisotropic and chiral interactions, it can be expected that the twist in segment packing alters the shape of the composition profiles themselves away from the intrinsic, or isotropic, SCF model predictions~\cite{Zhao2013, Yang2022}.

\section{Conclusion}

In summary, we have shown that chirality transfer in alternating network phases of BCPs falls into an unusual category. The preference for handed rotation at the molecular level (backbone orientation of chiral blocks) is a reverse sense of the thermodynamically selected mesoscopic twist of the cubic network (inter-node dihedral). The scenario relies on the fact that the twist of both the polar order parameter {\it and} the principal director of the nematic order parameter are screened by strong SmC-like anchoring to the scalar composition gradients. The residual, dominating twist, therefore, emerges from the strongly biaxial texture of segmental order in BCP gyroids, and its specific chiral gradients.  Notably this ``anti-sense" coupling to mesoscopic chirality of the network is also distinct from recently explored mechanisms of mesogenic packing and chirality transfer in gyroidal liquid crystal phases~\cite{Chen2020, Cao2020, Dressel2020}, wherein the principal director is understood to co-rotate with the ``easy'' 70.5$^\circ$ inter-nodal twist shown in Fig.~\ref{fig:aDG}.  

We also note that this mechanism of chirality transfer is distinct from previously studied mechanisms for BCP*s, notably for H* and DG phases of chiral diblocks~\cite{Zhao2013, Yang2022}.  In those cases, in the absence of segmental chirality, the equilibrium phases are achiral. To overcome the entropic coupling that locks the director to composition gradients, the free energy gain for threading chiral segment packing must be sufficiently strong. This leads to a threshold value of $q_0$ below which the structure remains mesoscopically achiral; and above which it breaks inversion symmetry.  In this case, equilibrium alternating networks of ABC triblocks spontaneously break chiral symmetry. The preference for chiral segment packing needs to only tip the balance between otherwise thermodynamically degenerate enantiomeric phases. Hence, at the mean-field level, chirality transfer occurs in the limit of arbitrarily weak chiral strength. It remains to be studied, how the domain shapes of alternating networks are modified in the presence of strong nematic chiral couplings, which could, in principle, ``unwind" the handedness reversal effect studied here for weak chirality and would be consistent with the twist preferences of (polar) strongly chiral diblocks in the DG assemblies~\cite{Yang2022}.  If that were the case, it would suggest an even more complex scenario, where the left vs. right. handedness selection of network formed by the chiral block in alternating gyroids is itself a function of the chiral strength, switching from anti-chiral to homo-chiral transfer when going from weakly to strongly chiral BCP* systems.

\section{Acknowledgements}
The authors are grateful to E. Thomas, R.-M. Ho and R. Kamien for valuable discussions regarding this work, and further to C. Burke for input on resolution limits of intra-domain twist.  This research was supported by US AFOSR under Asian Office of Aerospace Research and Development Award 18IOA088.  Simulations where performed using the UMass Cluster at the Massachusetts Green High Performance Computing Center.

\appendix

\section{Orientational self-consistent field theory}
\label{sec: oSCF}

Here we summarize the key theoretical ingredients for the oSCF theories of ABC triblocks with generalized couplings between freely-jointed chain random walks and either {\it polar} or {\it nematic} orientational field interactions.  The complete details and derivations of this formalism can be found in refs. \cite{Zhao2012,Prasad2017, Burke2018}.  Here, we describe only elements necessary to analyze chirality transfer in the limit of {\it weakly anisotropic} couplings. 

In general, we consider mean-field interactions of the form described by local functions of the order parameters. For example, the standard (Flory-Huggins) model of chain mixtures includes an enthalpic free energy of interactions between unlike components in terms of overlap between {\it scalar} composition profiles, $\phi_\alpha (\xv)$, which for the case of ABC copolymers, this takes the form,
\begin{multline}
    F_{\rm scal} \big[ \phi_\alpha (\xv) \big] = \rho_0 \int d^3 \xv \Big\{ \chi_{AB}\phi_A \phi_B  +\chi_{BC}\phi_B \phi_C \\ +\chi_{CA}\phi_C \phi_A \Big\} .
\end{multline}
Additionally, we consider the possibility of orientational interactions described by functionals of either vector (polar) or tensorial (nematic) form, $F_{\rm pol} \big[ {\bf p}_\alpha (\xv) \big]$ and $F_{\rm nem} \big[ {\bf Q}_\alpha (\xv) \big]$, respectively. Where these functionals depend on local values of the order parameters and their gradients. These orientational couplings give rise to self-consistent vectorial and tensorial fields, which are denoted as ${\bf w}_\alpha$ and ${\bf W}_\alpha$ respectively, representing the anisotropic interactions between chain segments at the mean-field level, and derive from the saddle point equations:
\begin{equation}
    w_{\alpha, i} ({\bf x} ) = \rho_0 \frac{ \delta F_{\rm pol}} { \delta p_{\alpha,i} ({\bf x} ) },
\end{equation}
and 
\begin{equation}
\label{eq: tensorial}
    W_{\alpha, ij} ({\bf x} ) = \rho_0 \frac{ \delta F_{\rm nem}} { \delta Q_{\alpha,ij} ({\bf x} ) },
\end{equation}
where index $\alpha$ refers to chemical component (i.e. A, B, or C type) and $i$ and $j$ are spatial indices.  These anisotropic interaction fields bias the random-walk chain statistics through diffusion-like equations for the segment distributions, $q^{\pm}({\bf x}, n)$, given respectively by, 
\begin{equation}
   \pm \frac{\partial q^{\pm}}{\partial n} = \frac{1}{6}\big[a \nabla -{\bf w}_\alpha ({\xv})\big]^2 q^{\pm} - \omega_{\alpha}({\bf x} ) q^{\pm}, \ \ \ \ \ {\rm (polar)} 
\end{equation}
and 
\begin{equation}
   \pm \frac{\partial q^{\pm}}{\partial n} = \frac{a^2}{6} \nabla^2 q^{\pm} -\frac{a^2}{15} \partial_i \big[ \tilde{W}_{\alpha, ij}({\bf x}) \partial_j q \big]  - \omega_{\alpha}({\bf x} ) q^{\pm}, \ \ \ \ \ {\rm (nematic)} ,
\end{equation}
where $\tilde{W}_{ij} = (W_{ij}-W_{ji})/2- (\delta_{ij}/3)~ {\rm Tr}  [ {\bf W}]$ is the traceless, symmetric part of $W_{ij}$ and $\omega_{\alpha}({\bf x} )$ is the self-consistent chemical potential fields that couple to the scalar composition profiles for component $\alpha$.  The self-consistency conditions for the  $\omega_\alpha ({\bf x})$ fields follow from the standard saddle-point conditions relating to the functional derivatives $\delta F_{\rm scal}/\delta \phi_\alpha ({\bf x})$ and the incompressibility constraint, $\sum_{\alpha} \phi_\alpha ({\bf x} ) =1 $.  As the anisotropic self-consistent fields bias the random walk statistics of chains, they couple to the mean-field orientational order parameters, providing a complete set of self-consistency conditions for the generalization of block copolymer melts to the case of anisotropic interactions:
\begin{equation}
\label{eq:polar}
    {\bf p}_\alpha ({\bf x}) = \frac{V}{6 N{\cal Z}} \int_{n \in \alpha} dn~ \Big[ a {\bf J} (\xv, n) - 2 {\bf w} ({\bf x}) q^+ q^-\Big] , \ \ \ \ \ {\rm (polar)} 
\end{equation}
and 
\begin{multline}
\label{eq:nematic}
    {\bf Q}_\alpha (\xv) = \frac{V}{N {\cal Z}} \int_{n\in \alpha} dn \bigg\{a^2 \Big( {\bf K} (\xv, n) - \frac{{\bf I}}{3} {\rm Tr}\big[{\bf K}(\xv, n)\big] \Big) \\ - 8 \tilde{\bf W}({\bf x}) q^+ q^-\bigg\}, \ \ \ \ \ {\rm (nematic)} 
\end{multline}
where ${\bf J} (\xv, n)$ and ${\bf K} (\xv, n)$ are vectorial and tensorial segment-flux operators defined in the main text.

In the limit where the anisotropy goes to zero (i.e. $F_{\rm pol} \to 0$ and $F_{\rm nem} \to 0$), the self-consistent vectorial and tensorial fields vanish.  In this case,  ${\bf p}_\alpha$ $ {\bf Q}_\alpha$ derive only from spatial derivatives of $q^{\pm}(n,\xv)$, corresponding to the form of the ``intrinsic'' polar and nematic order parameters given in the main text eqs. (2) and (3).

We note that these intrinsic order parameters have an explicit dependence on $N$ which can be deduced by considering the natural length scales of the microphase separation. The characteristic size of BCP domains scale with the r.m.s. size of an unperturbed Gaussian chain, $N^{1/2}a$, with an additional dependence on values of $\chi_{\alpha \beta} N$ and $f_{\alpha}$.  This implies that all spatial variation in $q^{\pm}$ occurs on a length scale that grows as $N^{1/2}a$.  Hence, spatial derivatives of the form that enter eqs. (2) and (3) in the main text are made dimensionless by multiplication by $N^{1/2}a$ (i.e. $\bar{\partial}_i \to N^{1/2}a \partial_i$).  Recasting these derivative conditions in terms of reduced coordinates, it is clear that $N^{1/2} {\bf p}_\alpha$ and  $N {\bf Q}_\alpha$ are functions that only depend on $\chi_{\alpha \beta} N$ and $f_{\alpha}$.

\section{Intrinsic order parameter and twist}
\label{sec: intrinsic}

To evaluate the intrinsic order parameters, and the sub-domain patterns of twist, we compute the chain distribution functions $q^{\pm}(n,\xv)$ for ABC triblocks using the PSCF code~\cite{Arora2016}.  We then compute the polar and nematic order parameters using eqs. (\ref{eq:polar}) and (\ref{eq:nematic}), and a finite-difference approximation to the spatial derivatives of $q^{\pm}(n,\xv)$ defined on the computational grid of the SCF solution.  

\begin{figure*}
% \centering\includegraphics[width=1.0\linewidth]{fig_monomer_design.eps} 
\centering\includegraphics[width=0.85\linewidth]{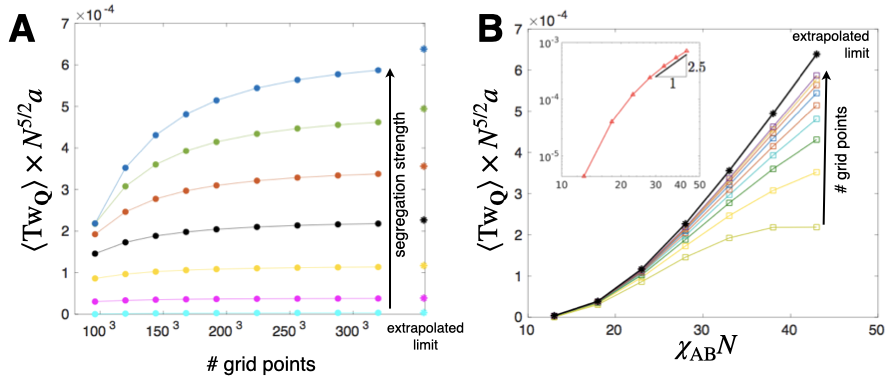} 
\caption{ In (A) we plot the mean nematic twist in the A network domain for SCF grids of variable grid resolution, and a sequence of segregation strengths given by the points in (B).  In (B) we plot the mean twist as a function of segregation strength, for the sequence of increasing grid resolutions plotted in (A).  The extrapolated limit is shown on a log-log plot on the inset, suggesting a power law scaling of $(\chi_{AB} N)^{5/2}$.  Here, we consider the case of $f_A=f_C = 0.24$}
\label{fig:twistresolution}
\end{figure*}

Polar and nematic order parameters require spatial derivatives of $q^{\pm}(n,\xv)$ and the resolution of segment twist is another spatial derivative of the order parameter. Thus, converged oSCF solutions require especially high grid resolution, even beyond the demands for the free energy of standard SCF theory (i.e. isotropic Flory interactions)~\cite{Burke2018}. To determine the resolution limits of twist calculations we analyze the nematic twist (which requires higher resolution than polar twist) as a function of the grid resolution in Fig. ~\ref{fig:twistresolution}. We observe that nematic twist eventually converges with sufficiently large numbers of spatial grid points, in a way that increases with the segregation strength.  Fitting that limiting convergence to an asymptotic power law, we extrapolate to limiting values in the hypothetical infinite grid resolution limit. Results for Fig.~\ref{fig:intradomain}C are based on $320^3$ grid elements, while the phase diagrams, at much lower segregation strengths, are based on $192^3$ spatial grids.

\begin{figure}
% \centering\includegraphics[width=1.0\linewidth]{fig_monomer_design.eps} 
\centering\includegraphics[width=1.0\linewidth]{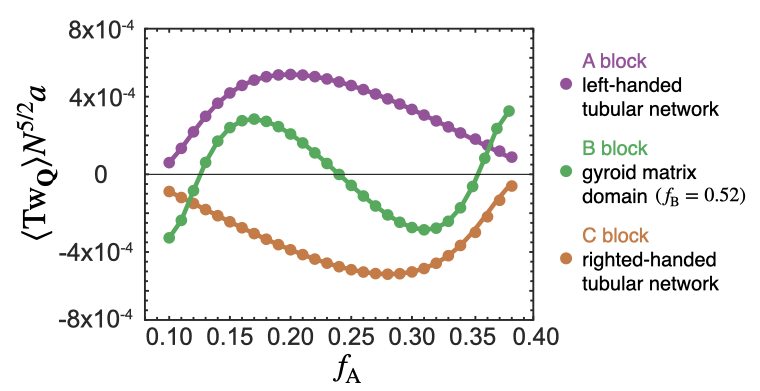} 
\caption{ A plot of mean nematic twist of all three domains for $\chi_{AB} N = \chi_{BC} N= \chi_{AC} N/2.7=41$ as function of end-block fraction $f_A$ for a fixed midblock fraction $f_B=0.52$.  Note that end block compositions are symmetric when $f_A=0.24$.}
\label{fig: matrix}
\end{figure}

In Fig.~\ref{fig: matrix}, we consider the mean nematic twist in all three domains of an aDG phase for the variable asymmetry between the end blocks for a fixed midblock fraction $f_B=0.52$.  This shows that when $f_A \neq f_C$ the matrix itself also develops a net chirality in segment twist.  Notably, the decreasing length relative length one of the two ends (say, the A block) leads to a transfer to the chirality of twist from the tubular domain to the matrix.  For example, when $f_A < f_C$ the net twist of the matrix layer has the sign of the A domain twist, and this behavior inverts when $f_A > f_C$.

\section{Alignment of polar and nematic directors with composition gradients}

\label{sec: aligment}

\begin{figure*}
% \centering\includegraphics[width=1.0\linewidth]{fig_monomer_design.eps} 
\centering\includegraphics[width=0.7\linewidth]{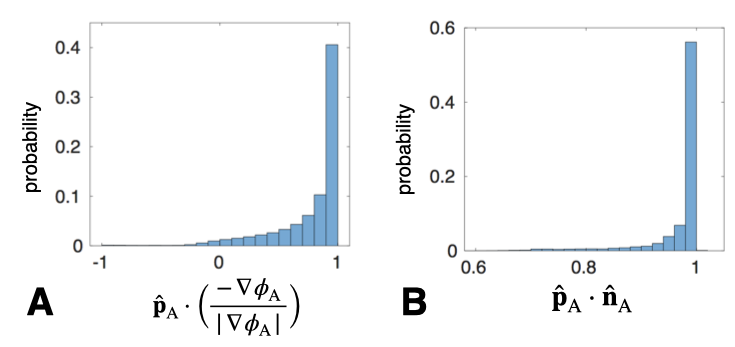} 
\caption{ Histograms of the relative alignment of isodensity contours and polar order (A) and polar and nematic order (B) for the single-gyroid A domain of an aDG phase.  Here we analyze the same structure shown in Fig.\ref{fig:intradomain}A.}
\label{fig:alignment}
\end{figure*}

Ref. ~\cite{Prasad2017} described the general ``anatomy" of the segmental order parameters of BCP domains in detail.  Here we focus on the relative alignment between polar order parameter and isocontours of segment composition for the A domain, composing a single gyroid domain of aDG.  The isocontours of density are normal to $-\nabla \phi_A/|\nabla \phi_A|$, and in Fig.~\ref{fig:alignment}A we plot the histogram of the dot product of $ {\bf \hat{p}}_A$ and these isocontour normals, showing that these directions are predominantly parallel.  The principal director of nematic order is in general more complex due to a strong tangential alignment effect that takes place at the IMDS.  However, as the interface between network A and matrix B is very narrow is well-segregated morphologies, this tangentially aligning zone makes up a small fraction of the morphology.  As a consequence polar, and principal directors of the nematic, are on average very well aligned, as shown in Fig.~\ref{fig:alignment}B.  

\section{Decomposition of biaxial twist}
\label{sec: biaxial}

We decompose the twist of  a biaxial nematic order parameter parameterized by a principal director, $\nv$, and biaxial director, $\mv$, which are mutually perpendicular (i.e. $\nv \cdot \mv=0$), and the $Q_{ij}$ tensor,
\begin{equation}
Q_{ij} =Q^n_{ij}+  Q^m_{ij} ,
\end{equation}
where,
\begin{equation}
Q^n_{ij}= S \big( n_i n_j - \frac{\delta_{ij}}{3} \big) ; \ Q^m_{ij}= R \big( m_i m_j - \frac{\delta_{ij}}{3} \big) ,
\end{equation}
where magnitudes $S$ and $R$, along with directors, may vary in space.  We can relate this description to the three principal eigenvalues of $Q_{ij}$,
\begin{equation}
\lambda_1 = \frac{1}{3} (2S - R) ; \ \lambda_2 = \frac{1}{3} (2R - S) ; \ \lambda_3 = -\frac{1}{3} (S + R) .
\end{equation} 
From this we compute the twist of the tensor OP,
\begin{equation}
{\bf Q} \cdot \big( \grad \times {\bf Q} \big ) = {\rm Tw}_n+{\rm Tw}_m+{\bf Q}^n \cdot \big( \grad \times {\bf Q}^m \big )+{\bf Q}^m \cdot \big( \grad \times {\bf Q}^n \big ) ,
\end{equation}
where $ {\rm Tw}_\mu = {\bf Q}^\mu \cdot \big( \grad \times {\bf Q}^\mu \big ) $ is simply the twist of the principal or secondary (biaxial) director, e.g.
\begin{multline}
 {\rm Tw}_n = Q^n_{ij} \epsilon_{jk\ell} \partial_k Q^n_{\ell i}  \\ =Q^n_{ij} \epsilon_{jk\ell} \Big[ (\partial_k S) \big( n_\ell n_i - \frac{\delta_{\ell i}}{3} \big)+ S\big(n_i \partial_k n_\ell+n_\ell \partial_k n_i\big) \Big] \\
  = \frac{S(\partial_k S) }{3}  \epsilon_{jk\ell} \big( n_\ell n_j + \frac{\delta_{\ell j}}{3} \big) + \frac{2S^2}{3} n_j  \epsilon_{jk\ell} \partial_k n_\ell \\ - \frac{S^2}{3} \epsilon_{jk\ell} n_\ell \partial_k n_j = S^2 \nv \cdot (\grad \times \nv )
 \end{multline}
 where we used the fact that terms that are symmetric in any two indices vanish when contracted with $\epsilon_{jk\ell} $.  Note that the index $\mu = n, m$ refers to the principal or secondary component of ${\bf Q}$.  Likewise, $ {\rm Tw}_m =R^2  \mv \cdot (\grad \times \mv )$.  Notice that gradients of the OP {\it magnitudes} do not generate {\it unaxial} twist.  
 
Now we examine the cross term, ${\bf Q}^m \cdot \big( \grad \times {\bf Q}^n \big )$, beginning with the curl tensor,
\begin{equation}
\big( \grad \times {\bf Q}^n \big )_{ij} = \epsilon_{jk\ell} \Big[ (\partial_k S) \big( n_\ell n_i - \frac{\delta_{\ell i}}{3} \big)+ S\big(n_i \partial_k n_\ell+n_\ell \partial_k n_i\big) \Big] ,
\end{equation}
When we contract the first term with $Q^m_{ij}$ we get
\begin{multline}
R (\partial_k S)   \epsilon_{jk\ell} \big( m_i m_j - \frac{\delta_{ij}}{3} \big)  \big( n_\ell n_i - \frac{\delta_{\ell i}}{3} \big) \\ = -\frac{ R (\partial_k S)  }{3} \epsilon_{jk\ell} (n_\ell n_j +m_\ell m_j ) = 0 
\end{multline}
where we have used $\nv \cdot \mv =0$ and the symmetry of $(n_\ell n_j +m_\ell m_j ) $ under the interchange of indices.  Now contracting $Q^m_{ij}$ with the second term in the curl tensor $\big( \grad \times {\bf Q}^n \big )_{ij} $
\begin{multline}
S R   \epsilon_{jk\ell} \big( m_i m_j - \frac{\delta_{ij}}{3} \big) n_i \partial_k n_\ell= -\frac{S R}{3}  \epsilon_{jk\ell} n_j\partial_k n_\ell \\ =-\frac{S R}{3} \nv \cdot (\grad \times \nv ) .
\end{multline}
Contracting $Q^m_{ij}$ with the third term in the curl tensor $\big( \grad \times {\bf Q}^n \big )_{ij} $ gives
\begin{multline}
S R   \epsilon_{jk\ell} \big( m_i m_j - \frac{\delta_{ij}}{3} \big) n_\ell \partial_k n_i \\ =S R \Big[ m_i (\partial_k n_i)    \epsilon_{jk\ell}  m_j n_\ell  - \frac{1}{3}    \epsilon_{jk\ell}   n_\ell \partial_k n_j \Big] \\ = SR \Big\{ \mv \cdot \big[ (\lv \cdot \grad) \nv \big] + \frac{1}{3} \nv \cdot (\grad \times \nv ) \Big\} ,
\end{multline}
where
\begin{equation}
\lv = \nv \times \mv ,
\end{equation}
is the eigendirection of the smallest eigenvalue of ${\bf Q}$.  Combining all three terms we have,
\begin{equation}
{\bf Q}^m \cdot \big( \grad \times {\bf Q}^n \big ) =SR~ \mv \cdot \big[ (\lv \cdot \grad) \nv \big] .
\end{equation}
For the other ``cross twist" term we have
\begin{multline}
{\bf Q}^n \cdot \big( \grad \times {\bf Q}^m \big ) =SR n_i (\partial_k m_i)    \epsilon_{jk\ell}  n_j m_\ell \\ = - SR \nv \cdot \big[ (\lv \cdot \grad) \mv \big] = ~SR \mv \cdot \big[ (\lv \cdot \grad) \nv \big]  ,
\end{multline} 
where we used the fact that $\partial_i (\mv \cdot \nv) =0$ so, $\nv \cdot \big( \partial_i \mv \big) = -\mv \cdot \big( \partial_i \nv \big)$. 

Thus, putting the pieces together we have three total terms in the OP twist,
\begin{equation}
{\bf Q} \cdot \big( \grad \times {\bf Q} \big ) = S^2  ~ \nv \cdot (\grad \times \nv )+R^2 ~ \mv \cdot (\grad \times \mv )+ 2SR ~  \mv \cdot \big[ (\lv \cdot \grad) \nv \big]  .
\end{equation}

\section{Mean field thermodynamics of chiral nematic ABC* triblock melts: weakly-anisotropic limit}

\label{sec:weak}

Here we derive the weakly-anisotropic limit of the chiral nematic model of ABC triblocks described in the main text, summarized by the gradient free energy expression in eq. (\ref{eq:FQ}) (i.e. with chiral anisotropic mean-field interactions in A block only). Following the eq. (\ref{eq: tensorial}), we have the self-consistent tensorial field equation
\begin{multline}
\label{eq: Wform}
    W_{A, ij} ({\bf x} ) = \rho_0 \frac{ \delta F_{\rm nem}} { \delta Q_{A, ij} ({\bf x} ) } \\ = \rho_0 K \Big\{-\frac{1}{4} \nabla^2 ({\bf Q}_{A})_{ij}+2 q_0 (\nabla \times {\bf Q}_{A})_{ij}+q_0^2 ({\bf Q}_{A})_{ij} \Big\}
\end{multline}
where again we are considering the single constant approximation $K_0=K_1=K$.  

We now consider solutions of the self-consistent equations in the limit of weakly-anisotropic interactions, which is based on a thermodynamic perturbation expansion around the $K \to 0$ limit, which is the case of the standard (purely scalar) SCF theory~\cite{Burke2018}, referred to here as the ``intrinsic field'' limit.  Following standard perturbation theory approaches, we denote the values of self-consistent fields $g({\xv})$ as an expansion in powers of $K$,
\begin{equation}
   g({\xv}) = g^{(0)} ({\xv}) + K g^{(1)} ({\xv}) + K^2 g^{(2)} ({\xv})  \dots
\end{equation}
where $g^{(M)}({\xv})$ describes the $K^M$ order correction to the intrinsic field limit of the field  $g({\xv})$.  Based on this definition and the form of eq. (\ref{eq: Wform}), it is clear that intrinsic limit of tensorial self-consistent field vanishes, i.e. ${\bf W}_A^{(0)} = 0$, where as all other self-consistent field quantities remain non-zero in this limit, most notably the nematic tensor field itself (i.e. ${\bf Q}_A^{(0)}$ is given by equation (3) in the main text).  The total free energy then has the form,
\begin{multline}
\label{eq: Feq}
    F = F^{(0)} + \frac{K}{4}  \int d^3 x \bigg\{ K_0 \big[ (\nabla \cdot {\bf Q}^{(0)}_A)_i \big]^2 \\ + K_1  \big[(\nabla \times {\bf Q}_A^{(0)})_{ij}+2 q_0 ({\bf Q}_A^{(0)})_{ij} \big]^2\bigg\} + {\cal O}(K^2)
\end{multline}
with the second term corresponding to the lowest order non-vanishing correction to the intrinsic field limit due to anisotropic segment interactions.  Note that because the intrinsic limit is a saddle point with respect to the scalar fields, corrections to the free energy due to first-order corrections in $\phi_\alpha$ and $\omega_\alpha$ lead to corrections of the mean-field free energy at ${\cal O}(K^2)$. Therefore, in the limit of $K \to 0$, we consider only the first-order contribution of anisotropy, as in eq. (\ref{eq: Feq}), deriving from the chiral nematic gradient free energy of the intrinsic field ${\bf Q}^{(0)}_A$.

To evaluate the free energies of competing morphologies in Fig.~\ref{fig:phase}, we compute the intrinsic limit SCF solutions for competing phases for a variable range of domain periodicities centered around the equilibrium state (i.e. in the absence of anisotropic interactions).  We then evaluate the nematic order parameters of the A block and compute the gradient free energy contributions from eq. (\ref{eq:FQ}).  Combining these with the isotropic mean free energies in eq. (\ref{eq: Feq}), we minimize the total free energy over the domain periodicities, which may shift the equilibrium domain spacing by a few percent.

\bibliography{chiralgyroid}% Produces the bibliography via BibTeX.

\end{document}